\documentclass[11pt,aps,prd ,tightenlines,nofootinbib,longbibliography, superscriptaddress,notitlepage]{revtex4-1}
\usepackage{xspace}
\usepackage[english]{babel}
\usepackage{amsfonts}
\usepackage{amsmath, amsthm, amssymb, mathrsfs}
\usepackage[dvips]{graphicx}
\usepackage[left=2cm, right=2cm, top=2.5cm, bottom=2cm]{geometry}
\usepackage{bbm}
\usepackage[pdfencoding=auto,psdextra]{hyperref}
\usepackage{enumerate}
\usepackage{color}
\usepackage{soul}

\usepackage{pstricks}
\usepackage{pstricks-add}
\usepackage{epsfig}
\usepackage{pst-grad}
\usepackage{pst-plot}

\newcommand{\abs}[1]{\left|#1\right|}

\newcommand{\ket}[1]{|#1\rangle}
\newcommand{\norm}[1]{\left|#1\right|}
\newcommand{\mean}[1]{\big\langle #1 \big\rangle}

\newcommand{\makeSymbol}[1]{\mathord{\vcenter{\hbox{#1}}}}

\begin{document}

\title{On the polymer quantization of connection theories:\\ graph coherent states}

\author{Mehdi Assanioussi}
\email[]{mehdi.assanioussi@fuw.edu.pl}
\email[]{mehdi.assanioussi@desy.de}
\affiliation{II. Institute for Theoretical Physics, University of Hamburg,\\ Luruper Chaussee 149, 22761 Hamburg, Germany.}
\affiliation{Faculty of Physics, University of Warsaw, Pasteura 5, 02-093 Warsaw, Poland.}

\begin{abstract}
We present the construction of a new family of coherent states for quantum theories of connections obtained following the polymer quantization. The realization of these coherent states is based on the notion of graph change, in particular the one induced by the quantum dynamics in Yang-Mills and gravity quantum theories. Using a Fock-like canonical structure that we introduce, we derive the new coherent states that we call the graph coherent states. These states take the form of an infinite superposition of basis network states with different graphs. We further discuss the properties of such states and certain extensions of the proposed construction.
\end{abstract}

\maketitle

\section{Introduction}

The polymer quantization is a quantization procedure developed in the context of background independent approaches to quantum field theories. The idea of background independence arises from classical theories of gravity, where it is understood that gravity is geometry encoded in a metric. Carrying this notion to the quantum realm requires giving up the standard Fock quantization, and reformulating theories in terms of variables (holonomies) which do not rely on any specific background geometry. The polymer quantization was first developed for gravity and gauge theories \cite{Polym1, Polym2, Polym4}, which eventually gave rise to the loop quantum gravity approach \cite{LQG0, LQG3, LQG2, LQG1, LQG4}. The loop quantum gravity program provides a framework where gravity and the fields of the Standard Model are quantized in a background independent fashion. The appellation polymer quantization comes from the fact that the basic quantum excitations of gravity and gauge fields manifest as one dimensional excitations, that is polymer-like excitations.

In the Hamiltonian formulation of background independent theories, the dynamics is encoded in constraints. In this context, the polymer quantization is a canonical quantization where the constraints must be implemented and solved in the quantum theory. Difficulties arise however, in the implementation and solving of the so called quantum scalar or Hamiltonian constraint. Beside the ambiguities that are encountered in the implementation, the usual difficulty in solving {\it a} quantum Hamiltonian constraint is related to the complexity of the action of the Hamiltonian operator on the kinematical Hilbert space. In fact, in case of gravity and gauge theories, the full kernel of the Hamiltonian operator, which would eventually form the physical Hilbert space, is yet unknown. This translates into several obstacles in the investigation and understanding of the quantum dynamics as well as the extraction of physical predictions from the theory. In such situation, it seems that other methods such as approximation schemes, gauge fixing \cite{Bod1, Bod2} or deparametrization \cite{KucTor1,KucTor2,RovSmo94,KucharRomano,BroKuc,GieThiem12}, provide more manageable frameworks, though not complete, where certain quantum gravity aspects can be studied.

There are several approximation schemes that one could rely on. Namely, perturbative methods \cite{ALM}, semi-classical analysis \cite{GT}, coherent states \cite{CS1, CS2, CS3, CS4, CS5, LS, ADLMS, ELM}, effective dynamics for restricted sectors \cite{DL}, or controlled truncations of the quantum degrees of freedom \cite{QRLG1, QRLG2, QRLG3}. Each of these approaches contributed significantly in improving the understanding of the dynamics and the symmetric sectors in polymer quantum theories. Deparametrization on the other hand allows to fully quantize a sector of the classical phase space and obtain a quantum theory with a physical Hilbert space and a physical Hamiltonian. The physical Hamiltonian operator generates evolution with respect to a matter clock, and the question of solving the constraint equation is entirely avoided. Nevertheless it remains that any physical prediction in a deparametrized theory relies on the choice of interesting physical states. In the case where the Hamiltonian operator is graph changing\footnote{For readers unfamiliar with the concept, it is explained in section \ref{S3}}, one would need states whose whatever desired properties do not get spoiled by the graph changing character of the Hamiltonian. It is within these perspectives that the work exposed in the present article has been developed. The subject of this article is the introduction of new coherent states for polymer quantum theories of connections with compact internal gauge group, which include gravity and Yang-Mills fields. We call these new states graph coherent states, as they exhibit a certain compatibility with a particular graph change, and they take the form of a superposition of basis states with different graphs.

The organization of the article is as follows: in the second section we present some preliminaries concerning polymer quantum theories of connections and the dynamics of gravity and Yang-Mills theories. Then in the third section we develop our construction of graph coherent states. We start with the general setup, then we detail the construction in the simplest example of Maxwell theory, and later we present the full and general construction. In the fourth section we discuss certain aspects of the construction and possible generalizations and interpretations. Finally we conclude with a summary and some outlooks.

% \clearpage
\section{Preliminaries}

\subsection{Kinematics of quantum theories of connections}

The polymer canonical quantization \cite{Polym4, LQG0, LQG3, LQG2, LQG1, LQG4} of a connection theory in four spacetime dimensions with a compact gauge group $G$ leads to a kinematical Hilbert space 
\begin{align}\label{Hdef}
 {\cal H}:=L^2 ({\cal A}_{G} , d\mu_{AL,G} )\ ,
\end{align}
that is the space of square integrable functions on the configuration space ${\cal A}_{G}$ of $G$ connections with the Ashtekar-Lewandowski measure $\mu_{AL,G}$ \cite{AL-measure}. The space ${\cal H}$ is isomorphic to the completion of the space of cylindrical functions, on the space of $G$ connections ${\cal A}_{G}$, with respect to the inner product defined by the Ashtekar-Lewandowski measure \cite{LQG3}. 

The space ${\cal H}$ admits a basis whose elements we call $G$-colored networks (the spin networks in the case of gravity). A $G$-colored network function is a function labeled by a (cylindrical equivalence class of an oriented embedded) graph $\Gamma$, a set of irreducible representations of the group $G$ (excluding the trivial representations) assigned to the edges of $\Gamma$, and a set of $G$ tensors assigned to the vertices. The space $\cal{H}$ can be then decomposed as an orthogonal sum
  \begin{align}\label{decomp} 
  {\cal{H}} =\bigoplus_{\Gamma} {\cal H}^{\Gamma}\ ,
  \end{align}
where $\Gamma$ ranges over all cylindrical equivalence classes of non oriented graphs \cite{LQG3}, and ${\cal H}^{\Gamma}$ is the Hilbert space spanned by the $G$-colored networks with graph $\Gamma$.

As we mentioned in the introduction, since the quantization is background independent, the canonical dynamics of the theory is encoded in constraints. In the context of connection theories there are three constraints: the Gauss constraint imposing invariance with respect to the local gauge group transformations, the vector or spatial diffeomorphism constraint imposing invariance with respect to spatial diffeomorphisms, and the Hamiltonian constraint generating time gauge transformations. The first two are implemented and solved in the quantum theory through group averaging procedures \cite{Polym4}, while the later can be implemented as an operator, but the quantum constraint equation it defines is difficult to solve due to the complicated action of the Hamiltonian operator on a $G$-colored network function. In the context of Yang-Mills and Einstein gravity theories, an aspect of this action is the graph change, that is, given a $G$-colored network function, the Hamiltonian operator maps this state to a superposition of $G$-colored network functions with different graphs. In the following, we expose the details of the Hamiltonian operators in Yang-Mills and Einstein gravity theories, and we focus on a specific proposal \cite{LQGSC} which induces a particular graph change. This particular graph change is the one we use in the construction of the graph coherent states in section \ref{S3}.
% \clearpage

\subsection{The quantum dynamics of Yang-Mills and Einstein gravity theories}

In the case of Yang-Mills coupled to Einstein gravity theories, the Hamiltonian constraint $H$ takes the form \cite{QSD5}
\begin{align}\label{HamConst}
 H(N):=\int_\Sigma d^3x\ N\left[ \frac{s}{2k\beta^2}\left( \frac{\epsilon_{ijk}\underline E^a_i\underline E^b_j \underline F_{ab}^k}{\sqrt{q}} + \left(1-s \beta^2\right) \sqrt{q} R \right) + \frac{q_{ab}}{2g^2\sqrt{q}} \left( E_i^a E_i^b + B_i^a B_i^b \right)\right]\ ,
\end{align}
where $\Sigma$ is the space-like hypersurface, $N$ is the lapse function, $s$ is the spacetime signature, $\kappa=8\pi G_N$ with $G_N$ being Newton's constant, $\beta$ is the Immirzi-Barbero parameter, $\underline E$ is the gravitational densitized triad conjugate to the Ashtekar-Barbero $SU(2)$ connection $\underline A$, $\underline F$ is the curvature of $\underline A$, $q$ is the determinant of the three metric $q_{ab}$ on $\Sigma$, $g$ is the coupling constant of the Yang-Mills field, $E$ is the electric field conjugate to the Yang-Mills vector potential $A$, and $B_i^a=\frac{1}{2}\epsilon^{abc}F_{bc}^i$ with $F$ being the curvature of $A$.

The quantization of the Hamiltonian functional \eqref{HamConst} can be performed following a regularization procedure. There are couple of established regularizations of the Hamiltonian functional \cite{QSD1, QSD2, QSD5, Curv, AALM, LQGSC}, one of the main differences between them is the regularization of the curvatures of the connections of gravity and Yang-Mills fields. In this work, we consider the regularization proposed in \cite{AALM, LQGSC}, where the holonomy replacing the curvature of the connection is taken along a closed oriented loop at a vertex of a preexisting graph, which does not overlap with any edge of that graph. We call such loop a {\it special loop}. However, for the purposes of the construction of the new states we are presenting here, we modify slightly that regularization. More precisely, in order to attach the loop in a diffeomorphism invariant fashion, one uses a prescription which associates a special loop to each pair of edges \cite{AALM, LQGSC}.
We modify this prescription in the way that if two edges $e_I$ and $e_J$ belong to the same germ $[e_I]$ at a vertex $v$, then given a third independent edge $e_K$, the special loops associated to the pairs $(e_Ie_K)$ and $(e_Je_K)$ are diffeomorphically equivalent. This in particular implies a change in the tangentiality conditions proposed in \cite{AALM, LQGSC}, the new condition we choose simply states that a special loop $\alpha_{IJ}$ associated to a pair $(e_Ie_J)$ at a vertex $v$ is tangent to the two edges $e_I$ and $e_J$ up to the first order only. The end point of this modification is that the new prescription guarantees that the special loops with the same orientation, and associated to pairs of edges which belong to the same pair of germs, are all diffeomorphically equivalent.

The Hamiltonian operator of \cite{LQGSC} is defined not on the kinematical Hilbert space ${\cal H}:=\cal{H_G}\otimes \cal{H_M}$, $\cal{H_G}$ and $\cal{H_M}$ being the kinematical Hilbert spaces of gravity and Yang-Mills field respectively (each defined as in \eqref{Hdef}), but on the vertex Hilbert space \cite{Hvtx}. The vertex Hilbert space ${\cal H}^{\rm vtx}$ is the Hilbert space of partial solutions to the vector constraints. Namely, given two sub-spaces ${\cal H}_{\cal G}^{\Gamma} \subset \cal{H_G}$ and ${\cal H}_{\cal M}^{\Gamma} \subset \cal{H_M}$ obtained from the decomposition \eqref{decomp}, the elements of ${\cal H}^{\rm vtx}$ are obtained by averaging the elements of each of the sub-spaces ${\cal H}_{\cal G}^{\Gamma}$ and ${\cal H}_{\cal M}^{\Gamma}$ with respect to all smooth diffeomorphisms which act trivially in the set of vertices ${\rm Vert}(\Gamma)$ of the graph $\Gamma$. The scalar product is naturally induced from the space ${\cal H}$ through a rigging map \cite{Hvtx}. The vertex Hilbert space also decomposes into a tensor product of a gravity and matter Hilbert spaces
\begin{align}
  {\cal{H}^{\rm vtx}} &={\cal H}_{\cal G}^{{\rm vtx}}\otimes {\cal H}_{\cal M}^{{\rm vtx}}=\bigoplus_{[\underline \Gamma],[\Gamma]} {\cal H}_{\cal G}^{[\underline \Gamma]}\otimes {\cal H}_{\cal M}^{[\Gamma]}\ ,
\end{align}
where now $[\underline \Gamma]$ and $[\Gamma]$ stand for the equivalence classes of graphs defined with respect to the action of all smooth diffeomorphisms which act trivially in the sets of vertices ${\rm Vert}(\underline \Gamma)$ and ${\rm Vert}(\Gamma)$ respectively. In what follows we drop the brackets in the notation of those classes.
% \clearpage

The final expression of the Hamiltonian operator corresponding to the functional \eqref{HamConst} is given through its action on a network function $\Psi=\psi_{\cal G}^{\underline \Gamma} \otimes \psi_{\cal M}^\Gamma \in {\cal H}_{\cal G}^{\underline \Gamma}\otimes {\cal H}_{\cal M}^{\Gamma}$ as
\begin{align}\label{HamOp}
    \hat H(N) \Psi= \left(\sum \limits_{\substack{\underline v\in \underline \Gamma\\ v\in \Gamma}} N(\underline v) \hat H_{\cal G}^{\underline v}+N(v)\hat H_{\cal M}^v\right)\psi_{\cal G}^{\underline \Gamma} \otimes \psi_{\cal M}^\Gamma\ ,
\end{align}
where
\begin{align}\label{GravHam}
    \hat H_{\cal G}^v := \hat R(\underline P) + Q_{\cal G}(v) \sum \limits_{I,J,K} \epsilon^{IJK} \left(\text{Tr}_N^{(\underline l)}\left[\underline h_{\alpha_{IJ}}  \underline h_{s_K}  [\underline h_{s_K}^{-1} , \hat{V}(\underline P)]\right]+\text{Tr}_N^{(\underline l)}\left[\underline h_{\alpha_{IJ}}  \underline h_{s_K}  [\underline h_{s_K}^{ -1} , \hat{V}(\underline P)]\right]^\dagger\right)\ ,
\end{align}
and
\begin{align}\label{YMHam}
 \hat H_{\cal M}^v:=\frac{1}{2g^2} &\sum_{I,J} \hat \Theta_I \hat \Theta_J \left( P_{i,I} P_{i,J} + Q_{\cal M}(v) X_{k,I} X_{k,J}\right)\ ,
\end{align}
with
\begin{align}\label{Ord}
 X_{k,I}:= \sum \limits_{\substack{K,L}} \epsilon^{IKL} \left(\text{Tr}_N^{(l)}\left[\tau^k h_{\alpha_{KL}}\right]+ \text{Tr}_N^{(l)}\left[\tau^k h_{\alpha_{KL}}\right]^\dagger \right)\ .
\end{align}
The capital indices in the (ordered) sums run through all the edges meeting at the vertices $\underline v$ and $v$, and $\text{Tr}_N$ stands for the normalized trace ($\text{Tr}_N[\tau^i \tau^j]=\delta^{ij}$). $\hat R(\underline P)$ and $\hat V(\underline P)$ are the curvature \cite{Curv} and volume \cite{Vol} operators respectively and they both depend only on the gravity fluxes $\underline P$. $\hat \Theta_I$ are gravitational operators \cite{QSD5} acting exclusively on the space ${\cal H}_{\cal G}^{{\rm vtx}}$. $\underline h$ and $h$ are respectively gravity and Yang-Mills holonomy operators chosen in fixed representations\footnote{Since the construction of the Hamiltonian operator and its properties, as well as the construction of the coherent states we are presenting here do not depend on the specific choice of the representations of those holonomies, the representations are left arbitrary and are only assumed to be fixed.} labeled by $\underline l$ and $l$. $P$ are the Yang-Mills fluxes, $Q_{\cal G}(v)$ and $Q_{\cal M}(v)$ are determined factors which partially depend on the valence of the vertices, and finally $\epsilon^{IJK}=sgn(\text{det}[\dot e_I,\dot e_J,\dot e_K])$ where $\dot e$ stands for the tangent vector of the edge $e$ at the vertex $v$. In \eqref{GravHam} and \eqref{Ord}, we imposed a choice of ordering of the basic operators, and a choice of symmetrization of the summed terms using the adjoint operators denoted by $^\dagger$, it is the adjoint action on the space ${\cal H}_{v}^{\Gamma^{\cal A}}$, and it is not to be confused with the adjoint element in the group.

Due to the presence of the holonomy operators $\underline h_\alpha$ and $h_\alpha$, the operators $\hat H_{\cal G}^v$ and $\hat H_{\cal M}^v$ are graph changing. Meaning they map the graphs they act on to other graphs with a different distribution of special loops at the vertices. Schematically, their successive action on a given vertex $v$ of a graph $\Gamma$ gives
\begingroup
\allowdisplaybreaks
\begin{align}\label{Diagr}
\makeSymbol{\includegraphics[scale=1.7]{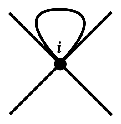}} \quad \longrightarrow \quad \makeSymbol{\includegraphics[scale=1.85]{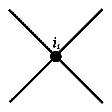}} + \makeSymbol{\includegraphics[scale=2]{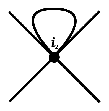}} + \makeSymbol{\includegraphics[scale=1.9]{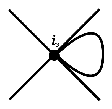}} + \makeSymbol{\includegraphics[scale=2]{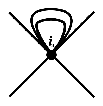}}+\quad \dots\quad .
\end{align}
\endgroup
Thanks to the fact that special loops associated to the same pair of germs are diffeomorphically equivalent, the special loops depicted on the fourth diagram on the right hand-side of \eqref{Diagr} are indistinguishable. They are indistinguishable in the sense that no physical observable could distinguish between special loops associated to the same pair of germs. This is a key property that we use in our construction of the graph coherent states.
% \clearpage

\section{Graph coherent states in quantum theories of connections}\label{S3}

In this section we will present a method to construct a family of coherent states in a generic Hilbert space $\cal{H}^\text{vtx}$ of a polymer quantum theory of connections with arbitrary compact gauge group $G$ (e.g.\ $\cal{H}_{\cal G}^\text{vtx}$ or $\cal{H}_{\cal M}^\text{vtx}$). These coherent states are constructed such that they take the form of a superposition of $G$-colored networks with different graphs. Though inspired from a particular dynamics, the construction is purely kinematical in the sense that it is realized on the Hilbert space $\cal{H}^\text{vtx}$ independently of the dynamics of the theory, and in principle it can be applied with a different graph change as we discuss briefly in section \ref{S4}.

Our construction is based on the observation that the action of the Hamiltonian, using the special loop regularization mentioned above, provides a decomposition of the Hilbert space $\cal{H}^\text{vtx}$ into separable subspaces, which are stable under the action of the Hamiltonian operator \eqref{HamOp}. As we will see in detail later, the separability of these subspaces induces an isomorphism between each of these subspaces and the Hilbert space of a given finite number of quantum harmonic oscillators. This isomorphism is a crucial step in order to obtain the canonical structure that we use to construct the graph coherent states.
Given one of these stable subspaces, one takes the colored graph\footnote{Definition: a colored graph is a graph with irreducible representations assigned to its edges.} of a single arbitrary element of the $G$-colored network basis spanning this space, and by removing all the special loops in this colored graph, one obtains what we call the {\it ancestor} graph. It then follows that the colored graphs of all the elements of the $G$-colored network basis in this space can be generated by attaching special loops at the vertices of the ancestor graph. Hence one can label the stable subspaces by the associated ancestor graphs $\{\Gamma^{\cal A}\}$, i.e.\ colored graphs with no special loops, and we denote them ${\cal H}^{\Gamma^{\cal A}}$. We then have
\begin{align}
{\cal H}^{\text{vtx}}=\bigoplus_{\Gamma^{\cal A}} {\cal H}^{\Gamma^{\cal A}}\ .
\end{align}
Furthermore, considering the local nature of the attachment of the special loops, i.e.\ it concerns each vertex of the graph separately, one can focus the analysis on a single vertex of a given ancestor graph, the generalization to the full graph is then straightforward. In other words, once an ancestor graph $\Gamma^{\cal A}$ is fixed, the only degrees of freedom left are the numbers of loops associated to the pairs of germs at each vertex, and the $G$ tensors at the vertices. Hence we can write
\begin{align}\label{Isom}
 {\cal H}^{\Gamma^{\cal A}} \cong \bigotimes_{\ v\in \Gamma^{\cal A}} {\cal H}_v^{\Gamma^{\cal A}}\ ,
\end{align}
meaning that the space ${\cal H}^{\Gamma^{\cal A}}$ for a given $\Gamma^{\cal A}$ is isomorphic to the tensor product of spaces ${\cal H}_v^{\Gamma^{\cal A}}$ each associated to a vertex $v$ of $\Gamma^{\cal A}$. The spaces ${\cal H}_v^{\Gamma^{\cal A}}$ are constructed as follows. Given an ancestor graph $\Gamma^{\cal A}$, each pair of germs $([e_I], [e_J])$ meeting at a vertex $v$ of $\Gamma^{\cal A}$ defines two oriented wedges $\omega_{IJ}^v$ and $\omega_{JI}^v$. The graph of any colored network in ${\cal H}^{\Gamma^{\cal A}}$ is given by the ancestor graph $\Gamma^{\cal A}$ and a number of special loops associated to the oriented wedges of $\Gamma^{\cal A}$. The special loops are oriented following the orientation of the oriented wedges they are associated to \cite{AALM, LQGSC}. 
Note, however, that depending on the gauge group $G$, the chosen operator to induce the graph change, and the graph $\Gamma^{\cal A}$, the states in ${\cal H}^{\Gamma^{\cal A}}$ which differ by the orientation of a special loop may span the same subspaces of ${\cal H}^{\Gamma^{\cal A}}$, either because of a specific relation between the operators creating the special loops with opposite orientations, or because of the symmetries of the graph $\Gamma^{\cal A}$. For instance, in Maxwell theory and in absence of symmetries of the graph, one can add special loops at a wedge $\omega_{IJ}^v$ by acting with the operator $\text{Tr}_N^{(l)}\left[\tau^k h_{\alpha_{IJ}}\right]$, present in the Hamiltonian operator \eqref{YMHam}. Because the gauge group is $U(1)$, we have that
\begin{align}
 \text{Tr}_N^{(l)}\left[\tau^k h_{\alpha_{IJ}}\right] = h_{\alpha_{IJ}}^{(l)}\ ,
\end{align}
and since $h_{\alpha_{JI}}^{(l)} = (h_{\alpha_{IJ}}^{(l)})^{-1}$, we conclude that the operators $\text{Tr}_N^{(l)}\left[\tau^k h_{\alpha_{IJ}}\right]$ and $\text{Tr}_N^{(l)}\left[\tau^k h_{\alpha_{JI}}\right]$ are linearly independent. Hence the oriented wedges $\omega_{IJ}^v$ and $\omega_{JI}^v$ are not equivalent, and each pair of germs provides two independent oriented wedges.

As we explain in section \ref{2B}, if the gauge group is $SU(2)$ and we choose the operator $\text{Tr}_N^{(l)}\left[\tau^k h_{\alpha}\right]$ to induce the graph change, then each pair of germs defines two equivalent oriented wedges from the perspective of special loops. In this case, it is enough to pick one orientation for the special loops and drop the second one. When it is present, this freedom in the choice of the independent wedge, associated to a pair of germs, is absorbed into the freedom of choosing a canonical structure associated to the pair of germs, we particularly illustrate this fact in the example of \eqref{Exp1}. 

We denote the set of all independent oriented wedges (which from now on we call simply wedges) at a vertex $v$ by ${\cal W}_v$ and its cardinality $w_v$.
To each vertex $v$ we associate the spaces $\{{\cal S}_{v,n}^{\Gamma^{\cal A}}\}_{n\in \mathbbm{N}}$, such that each ${\cal S}_{v,n}^{\Gamma^{\cal A}}$ is the space of states which describe the distribution of $n$ special loops at $v$, i.e.\ the association of $n$ special loops to the different wedges of $\Gamma^{\cal A}$ at $v$. Each space ${\cal S}_{v,n}^{\Gamma^{\cal A}}$ is spanned by an orthonormal basis whose elements are labeled by $w_v$ integers, which sum up to the total number $n$ and correspond to the numbers of special loops associated to the wedges of $\Gamma^{\cal A}$ at $v$. Then to each space ${\cal S}_{v,n}^{\Gamma^{\cal A}}$ is associated a space ${\cal I}_{v,n}^{\Gamma^{\cal A}}$ of admissible $G$ tensors, those are the $G$ tensors which couple the holonomies meeting at the vertex $v$. Finally, we define the space ${\cal H}_v^{\Gamma^{\cal A}}$ as
\begin{align}\label{DecGI}
{\cal H}_v^{\Gamma^{\cal A}}:= \bigoplus\limits_{n=0}^{\infty}\ {\cal H}_{v,n}^{\Gamma^{\cal A}} :=\bigoplus\limits_{n=0}^{\infty}\ {\cal S}_{v,n}^{\Gamma^{\cal A}}\otimes {\cal I}_{v,n}^{\Gamma^{\cal A}}\ .
\end{align}
It is then clear that the isomorphism in \eqref{Isom} holds.

Let us again point out that, thanks to the prescription of special loops in the regularization procedure, the special loops associated to a wedge of a graph are diffeomorphically equivalent, and hence indistinguishable from the perspective of physical observables.

For clarity we gradually develop the details of our construction: we first start with the simplest example of an abelian gauge group, namely Maxwell theory. Then we extend to the general case with arbitrary compact gauge group.

\subsection{Loop quantum Maxwell theory}

In Maxwell theory the internal gauge group is $U(1)$. Having an abelian gauge group implies that the only degrees of freedom left to characterize the basis states in the Hilbert space ${\cal H}^{\Gamma^{\cal A}}$ are the numbers of indistinguishable special loops associated to the wedges of $\Gamma^{\cal A}$. Using the decomposition \eqref{Isom}, one then has
\begin{align}
{\cal H}_{v}^{\Gamma^{\cal A}}\cong \bigoplus\limits_{n=0}^{\infty}\ {\cal S}_{v,n}^{\Gamma^{\cal A}}\ .
\end{align}
Knowing the structure of the spaces ${\cal S}_{v,n}^{\Gamma^{\cal A}}$, and using the indistinguishableness property of the special loops, it naturally follows that the space ${\cal H}_{v}^{\Gamma^{\cal A}}$ is isomorphic to the space of a multi-dimensional (or a finite number of) quantum harmonic oscillators, where to each wedge $\omega_i\in {\cal W}_v$ ($i\in \{1,\dots,w_v\}$) of $\Gamma^{\cal A}$ at $v$ is associated a space ${\cal F}_i$ of a quantum harmonic oscillator,
\begin{align}
{\cal H}_{v}^{\Gamma^{\cal A}}\cong \bigotimes_{i=1}^{w_v} {\cal F}_i\ .
\end{align}
In this picture, a wedge with a certain number of special loops corresponds to an energy level for a single quantum harmonic oscillator.

Given the spaces ${\cal F}_i$, we introduce a canonical structure on them through annihilation and creation operators $\{a_i,a^{\dagger}_i\}$ satisfying
\begin{align}
 \forall\ \omega_i, \omega_j \in {\cal W}_v,\quad [a_i,a_j]=[a^{\dagger}_i,a^{\dagger}_j]=0\quad,\quad [a_i,a^{\dagger}_j]=\delta_{ij}\mathbbm{I}\ .
\end{align}
The  vacuum state in each ${\cal F}_i$ is defined as
\begin{align}
a_i\ket{0_i}=0\ ,
\end{align}
and we take
\begin{align}
a_i\ket{n_i}=\sqrt{n_i}\ket{n_i-1}\quad,\quad a_i^\dagger \ket{n_i}=\sqrt{n_i+1}\ket{n_i+1} \ .
\end{align}
From the perspective of the Hilbert space ${\cal H}_{v}^{\Gamma^{\cal A}}$, the vacuum state in each ${\cal F}_i$ represents the wedge $\omega_i$ with no special loops, and the state $\ket{n_i}$ represents $\omega_i$ with $n$ special loops attached to it.

We then define the (normalized) canonical coherent states, i.e.\ eigenvectors of the annihilation operators:
\begin{align}
\forall\ \omega_i \in {\cal W}_v,\quad a_i\ket{z_i}=z_i\ket{z_i}\quad,\quad \ket{z_i} = e^{z_i a_i^\dagger-\bar{z}_i a_i} \ket{0_i}=e^{\frac{-\norm{z}^2}{2}}\sum_{n_i}\frac{z_i^{n_i}}{\sqrt{n_i!}}\ket{n_i}\quad,\quad z_i\in\mathbb{C} \ .
\end{align}
We call a state $\ket{z_i}$ a {\it graph coherent wedge} associated to the wedge $\omega_i$. We then introduce the coherent states $\{\ket{Z_v}\}$ as
\begin{align}\label{GCV}
\ket{Z_v}:=\bigotimes_{i=1}^{w_v}\ket{z_i}\quad,\quad Z_v:=\{ z_i\}\in\mathbb{C}^{w_v} \ ,
\end{align}
where each state $\ket{z_i}$ is a graph coherent wedge associated to the wedge $\omega_i$. We call the states $\{\ket{Z_v}\}_{Z_v\in\mathbb{C}^{w_v}}$ {\it graph coherent vertices} and they form an over-complete basis of the space ${\cal H}_{v}^{\Gamma^{\cal A}}$. By extension, through a tensor product over the vertices of ${\Gamma^{\cal A}}$, 
\begin{align}\label{GCS}
\ket{Z_{\Gamma^{\cal A}}}:=\bigotimes_{v \in \Gamma^{\cal A}}\ket{Z_v}\quad,\quad Z_{\Gamma^{\cal A}} \in \bigotimes \limits_{v\in \Gamma^{\cal A}}\mathbb{C}^{w_v} \ ,
\end{align}
one obtains an over-complete basis of {\it graph coherent states} $\{\ket{Z_{\Gamma^{\cal A}}}\}$ in the space ${\cal H}^{\Gamma^{\cal A}}$.

These graph coherent states are by construction coherent with respect to the standard combinations of sum and difference of the creation and annihilation operators. One could however investigate further the coherence properties of such states with respect to other operators of interest, namely the operators involved in the Hamiltonian operator $\hat H_{\cal M}^v$ \eqref{YMHam}. We first start by expressing these operators in terms of the annihilation and creation operators associated to the wedges of the ancestor graph at a vertex $v$. Simple calculations lead to the following identifications
\begin{align}
P_I &= j_I\ \mathbbm{I}\ ,\\ 
\text{Tr}_N^{(l)}\left[\tau h_{\alpha_{KL}}\right]&=h_{\alpha_{KL}}^{(l)}= a_{\tilde K\tilde L}^\dagger {\cal V}_{\tilde K\tilde L}\label{aToh}\ ,\\
\text{Tr}_N^{(l)}\left[\tau h_{\alpha_{KL}}\right]^\dagger&=[h_{\alpha_{KL}}^{(l)}]^\dagger={\cal V}_{\tilde K\tilde L}\ a_{\tilde K\tilde L}\ .\label{adaggerToh}
\end{align}
where each index $I$ selects an edge $I$ with color $j_I$ at the vertex $v$ of an arbitrary graph $\Gamma$ with $\Gamma^{\cal A}$ as ancestor graph, and a pair of indices $(KL)$ selects a wedge $\omega_{KL}$ of $\Gamma$ at $v$. In contrast, the tilded indices $\tilde K$, $\tilde L$ label the germs of the ancestor graph $\Gamma^{\cal A}$ at $v$, and the identification between $K$ and $\tilde K$ holds if and only if $e_K\in [e]_{\tilde K}$.
% $\dot e_K=\dot e_{\tilde K}$, that is $e_K$ and $e_{\tilde K}$ belong to the same germ $[K]=[\tilde K]$. 
The operator $\mathbbm{I}$ is the identity operator on the space ${\cal H}_{v}^{\Gamma^{\cal A}}$ and the operator ${\cal V}_{\tilde K\tilde L}$ is defined as
\begin{align}
{\cal V}_{\tilde K\tilde L}:=({\cal N}_{\tilde K\tilde L}+\mathbbm{I})^{-1/2}=(a_{\tilde K\tilde L}a_{\tilde K\tilde L}^\dagger)^{-1/2}\qquad ,\ \text{i.e.}\qquad {\cal V}_{\tilde K\tilde L}\ \ket{n}_{\tilde K\tilde L}= \frac{1}{\sqrt{n+1}}\ket{n}_{\tilde K\tilde L}\ ,
\end{align}
where ${\cal N}$ is the number operator. We remind the reader again that the dagger symbol $^\dagger$ in \eqref{adaggerToh} stands for the adjoint action on the space ${\cal H}_{v}^{\Gamma^{\cal A}}$, and it is not to be confused with the adjoint element in the group. Denoting by $\hat H_{\cal M}^{v,IJ}$ the operator
\begin{align}\label{YMHamFT}
\hat H_{\cal M}^{v,IJ}:=P_{I} P_{J} + Q_{\cal M}(v) X_I X_J\ ,
\end{align}
which is the part of the Yang-Mills Hamiltonian \eqref{YMHam} associated to a wedge of the graph $\Gamma$ at the vertex $v$, one obtains the following expression in terms of the new operators
\begin{align}\label{YMHamF}
 \hat H_{\cal M}^{v,IJ}= j_{I} j_{J}\mathbbm{I} +  Q_{\cal M}(v) \sum_{{\tilde M},{\tilde N},{\tilde K},{\tilde L}}\epsilon^{{\tilde I}{\tilde M}{\tilde N}}\epsilon^{{\tilde J}{\tilde K}{\tilde L}}\ \left(a_{\tilde M\tilde N}^\dagger {\cal V}_{\tilde M\tilde N}+{\cal V}_{\tilde M\tilde N} a_{\tilde M\tilde N}\right)\left(a_{\tilde K\tilde L}^\dagger {\cal V}_{\tilde K\tilde L}+{\cal V}_{\tilde K\tilde L} a_{\tilde K\tilde L}\right)\  ,
\end{align}
such that the pair $(e_Ie_J)$ belongs to the wedge $\omega_{\tilde I\tilde J}$.

Given the correspondence between the multiple operators in the expression of  $\hat H_{\cal M}^{v,IJ}$ and the canonical operators, we easily check the coherence properties of the graph coherent states defined in \eqref{GCS}. We explicitly compute the variance of these operators, these are given as follows (for clarity we drop the indices labeling the vertices and wedges)
\begingroup
\allowdisplaybreaks
\begin{align}
&\mean{(P_I P_J)^2}-\mean{P_IP_J}^2 =\mean{( {\cal V} a a^\dagger{\cal V})^2}-\mean{{\cal V} a a^\dagger{\cal V}}^2=0\label{Var0}\\
&\mean{( a^\dagger {\cal V} {\cal V} a)^2}-\mean{a^\dagger {\cal V} {\cal V} a}^2 =e^{-\norm{z}^2} \left(1-e^{-\norm{z}^2} \right)\\
&\mean{ ({\cal V} a)^2}-\mean{ {\cal V} a}^2 = z^2 e^{-\norm{z}^2}  \left(\sum_{n=0}^{\infty} \frac{\abs{z}^{2n}}{\sqrt{n!(n+2)!}}- e^{-\norm{z}^2} \left(\sum_{n=0}^{\infty} \frac{\abs{z}^{2n}}{\sqrt{n!(n+1)!}}\right)^2\right)\label{Var1}\\
&\mean{(a^\dagger {\cal V})^2}-\mean{a^\dagger {\cal V}}^2 =\bar z^2 e^{-\norm{z}^2}  \left(\sum_{n=0}^{\infty} \frac{\abs{z}^{2n}}{\sqrt{n!(n+2)!}}- e^{-\norm{z}^2} \left(\sum_{n=0}^{\infty} \frac{\abs{z}^{2n}}{\sqrt{n!(n+1)!}}\right)^2\right)\label{Var2}\ ,
\end{align}
\endgroup
and we estimate their relative variance $\Delta_r(.):=|\mean{(.)^2}/\mean{.}^2-1|$, obtaining
\begin{align}
\Delta_r({P_I P_J})=0\quad, \quad &\Delta_r({{\cal V} a a^\dagger{\cal V}})=0\quad, \quad
\Delta_r({a^\dagger {\cal V} {\cal V} a})=\frac{1}{e^{\norm{z}^2}-1}\ \xrightarrow[{\abs{z}\gg 1}]{} 0\ ,\label{RVar0}
\end{align}
and
\begin{align}
\Delta_r({ {\cal V} a})=\Delta_r({a^\dagger {\cal V}})&=1- \frac{e^{\norm{z}^2} \sum \limits_{n=0}^{\infty} \frac{\abs{z}^{2n}}{\sqrt{n!(n+2)!}}}{\left(\sum \limits_{n=0}^{\infty} \frac{\abs{z}^{2n}}{\sqrt{n!(n+1)!}}\right)^2}\ \xrightarrow[{\abs{z}\gg 1}]{} 0
\label{RVar2}\ .
\end{align}
Additionally, we consider the operator
\begin{align}\label{GermFlux}
 P_{\tilde I}:=\sum_{I\in \tilde I} |P_I|\ ,
\end{align}
where again $\tilde I$ stands for a germ of the ancestor graph. This operator could be interpreted as the (absolute) flux in the direction of the germ $\tilde I$. In terms of the canonical operators it becomes
\begin{align}
 P_{\tilde I}= \sum\limits_{I\in \tilde I} j_{I}\mathbbm{I}+ l \sum_{\tilde K} {\cal N}_{\tilde I \tilde K} + {\cal N}_{\tilde K \tilde I}\ ,
\end{align}
where $I$ runs through the edges of the ancestor graph belonging to the germ $\tilde I$, and $\tilde K$ runs through the germs of the ancestor graph meeting at the same vertex as $\tilde I$. The variance and the relative variance are estimated as
\begin{align}
&\mean{P_{\tilde I}^2}-\mean{P_{\tilde I}}^2= l^2\sum_{\tilde K} \abs{z_{\tilde I \tilde K}}^2 + \abs{z_{\tilde K \tilde I}}^2\quad,\quad \Delta_r(P_{\tilde I})\xrightarrow[{\abs{z_{\tilde I \tilde K}},\abs{z_{\tilde K \tilde I}}\gg 1}]{} 0\ .
\end{align}
The point of considering the operator $P_{\tilde I}$ is to show in an example that despite the fact that the canonical operators are directly linked to specific holonomy operators, one can construct operators depending only on fluxes and yet recover certain coherence properties with respect to the graph coherent states we introduced above. The reason why such properties may arise is that an operator such as $P_{\tilde I}$ describes a more global information about a given graph than just a flux operator associated to a single edge. In the case of $P_{\tilde I}$ this global information is captured by the abstract sum over the edges of the same germ, which translates into the appearance of the number operators in the expression of $P_{\tilde I}$, and thus exhibiting coherence properties with respect to the graph coherent states. Note that such abstract sums appear in operators which, for instance, are obtained from the quantization of non local (in space) functionals on the classical phase space, and they usually approximate the classical integrals over space-like regions. The Hamiltonian constraint and the volume of a space-like region are examples of such functionals, which are promoted to operators with abstract sums over the vertices and edges.
This fact sets the graph coherent states as promising states to probe the semi-classical properties of interesting physical observables through superposition of graphs.

In the following we present the general construction extending to arbitrary compact gauge group.

\subsection{Graph coherent states (II): beyond Abelian gauge groups}\label{2B}

The generalization of the above construction to a connection theory with a non abelian compact gauge group $G$ translates to taking into account the non trivial $G$ tensor spaces at the vertices of a colored graph. Namely, one has to incorporate the spaces ${\cal I}_{v,n}^{\Gamma^{\cal A}}$ in the implementation of a canonical structure on the spaces ${\cal H}^{\Gamma^{\cal A}}\subset{\cal H}^{\text{vtx}}$, and consequently in the definition of the graph coherent states.

\subsubsection{Generalized annihilation and creation operators}
Given a vertex $v$ of an ancestor graph ${\Gamma^{\cal A}}$, we consider a set of closed operators $\{a_i\}_{i\in {\cal W}_v}$ on ${\cal H}_v^{\Gamma^{\cal A}}$, each operator associated to a wedge $\omega_i$ at $v$, which satisfy
\begin{align}\label{OpInt0}
 \forall i\in {\cal W}_v,\quad a_i({\cal H}_{v,0}^{\Gamma^{\cal A}})=\{0\} \qquad \text{and}\qquad \forall n\geq 1,\quad a_i({\cal H}_{v,n}^{\Gamma^{\cal A}})\subset {\cal H}_{v,n-1}^{\Gamma^{\cal A}}\ .
\end{align}
In order for the set of operators $\{a_i\}$ and their adjoints $\{a_i^\dagger\}$ to form a canonical structure on the Hilbert space ${\cal H}_v^{\Gamma^{\cal A}}$, that is
\begin{align}\label{aCan}
 \forall i,j\in {\cal W}_v,\quad [a_i,a_j]=[a^{\dagger}_i,a^{\dagger}_j]=0\quad,\quad[a_i, {a_j}^\dagger]=\delta_{ij}\mathbb{I}_{{\cal H}_v^{\Gamma^{\cal A}}}\ ,
\end{align}
one must have a stronger condition than \eqref{OpInt0}, namely
\begin{align}\label{OpInt1}
 \forall i\in {\cal W}_v,\quad a_i({\cal H}_{v,0}^{\Gamma^{\cal A}})=\{0\} \qquad \text{and}\qquad \forall n\geq 1,\quad a_i({\cal H}_{v,n}^{\Gamma^{\cal A}})= {\cal H}_{v,n-1}^{\Gamma^{\cal A}}\ .
\end{align}
This implies that the operators ${a}_i^\dagger$ are injective maps on ${\cal H}_v^{\Gamma^{\cal A}}$ and,
since the spaces ${\cal H}_{v,n}^{\Gamma^{\cal A}}$ are of different dimensions, we have that
\begin{align}\label{Ker}
 \forall i\in {\cal W}_v,\quad \forall n\geq 1,\quad \dim\left(\text{Ker}[a_i]\cap {\cal H}_{v,n}^{\Gamma^{\cal A}}\right) = \dim\left({\cal H}_{v,n}^{\Gamma^{\cal A}}\right) - \dim\left({\cal H}_{v,n-1}^{\Gamma^{\cal A}}\right)\neq 0 \ .
\end{align}

We now provide in what follows a method to obtain such canonical structure. For simplicity, we focus the analysis on two wedges $\omega_1$ and $\omega_2$ at $v$, to each is associated a pair of operators $\{a_i,{a}_i^\dagger\}_{i=1,2}$ satisfying \eqref{OpInt1}. 
The construction is straightforwardly extendable to an arbitrary number of wedges at the vertex. Each space ${\cal H}_{v,n}^{\Gamma^{\cal A}}$ \eqref{DecGI} can be decomposed as
\begin{align}
 {\cal H}_{v,n}^{\Gamma^{\cal A}}=\bigoplus\limits_{n_1=0}^{n}\bigoplus\limits_{n_2=0}^{n-n_1}\ {\cal H}_{v,n_1,n_2}^{\Gamma^{\cal A}}\ .
\end{align}
where $n_1,n_2$ are the numbers of loops at the wedge $\omega_1$ and $\omega_2$ respectively. Denoting the normalized elements of ${\cal H}_{v,n_1,n_2}^{\Gamma^{\cal A}}$ by $\ket{\iota_{n_1,n_2}^\alpha}$, $\iota_{n_1,n_2}^\alpha \in {\cal I}_{v,n_1+n_2}^{\Gamma^{\cal A}}$, it follows from \eqref{OpInt1} that
\begin{align}\label{Cond1}
 \forall n_1,n_2 \in \mathbbm{N},\ &\forall \iota_{n_1,n_2}^\alpha\in {\cal I}_{v,n_1+n_2}^{\Gamma^{\cal A}},\ \exists !\ k_1\leq n_1,\ \exists !\ k_2\leq n_2\ :\\ \nonumber &\ a_1^{k_1} \ket{\iota_{n_1,n_2}^\alpha} \neq 0\ ,\ \forall p_1> k_1,\ a_1^{p_1} \ket{\iota_{n_1,n_2}^\alpha} = 0\ ,\\ \nonumber &\ a_2^{k_2} \ket{\iota_{n_1,n_2}^\alpha} \neq 0\ ,\ \forall p_2> k_2,\ a_2^{p_2} \ket{\iota_{n_1,n_2}^\alpha} = 0\ .&
\end{align}
We then denote our states as $\ket{\iota_{n_1,n_2;m_1,m_2}^\alpha}$ ($m_1\leq n_1$ and $m_2\leq n_2$) in order to encode the property \eqref{Cond1}, which is now expressed as
\begin{align}
 \forall n_1,n_2 \in \mathbbm{N},\ &\forall m_1\leq n_1,\ \forall m_2\leq n_2,\ \forall \iota_{n_1,n_2;m_1,m_2}^\alpha\in {\cal I}_{v,n_1+n_2}^{\Gamma^{\cal A}}\ :\\ \nonumber &\ a_1^{n_1-m_1} \ket{\iota_{n_1,n_2;m_1,m_2}^\alpha} \neq 0\ ,\ a_1^{n_1-m_1+1} \ket{\iota_{n_1,n_2;m_1,m_2}^\alpha} = 0\ ,\\ \nonumber &\ a_2^{n_2-m_2} \ket{\iota_{n_1,n_2;m_1,m_2}^\alpha} \neq 0\ ,\ a_2^{n_2-m_2+1} \ket{\iota_{n_1,n_2;m_1,m_2}^\alpha} = 0\ .
\end{align}
In other words, $m_1$ and $m_2$ denote the number of loops, at the wedges $\omega_1$ and $\omega_2$ respectively, in the vacuum state from which the state $\ket{\iota^\alpha_{n_1,n_2;m_1,m_2}}$ is obtained. We further discuss the notion of vacuum states in the part \ref{B2}.

We then choose to define the operators $\{a_i,{a}_i^\dagger\}$ through their actions on the states $\ket{\iota_{n_1,n_2;m_1,m_2}^\alpha}$ as follows
\begin{subequations}\label{GenAC2}
\begin{align}
 &\nonumber \forall n_1,n_2 \in \mathbbm{N},\ \forall m_1\leq n_1,\ \forall m_2\leq n_2,\ \forall \iota_{n_1,n_2;m_1,m_2}^\alpha \in {\cal I}_{v,n_1+n_2}^{\Gamma^{\cal A}}\ :\\ &\ a_1\ket{\iota_{n_1,n_2;m_1,m_2}^\alpha}=\sqrt{n_1-m_1}\  \ket{\iota_{n_1-1,n_2;m_1,m_2}^\beta}\ ,\\ 
 &\ {a}_1^\dagger\ket{\iota_{n_1,n_2;m_1,m_2}^\alpha}=\sqrt{n_1-m_1+1}\  \ket{\iota_{n_1+1,n_2;m_1,m_2}^\gamma}\ ,\\ 
 &\ a_2\ket{\iota_{n_1,n_2;m_1,m_2}^\alpha}=\sqrt{n_2-m_2}\  \ket{\iota_{n+1,n_2-1;m_1,m_2}^\delta}\ ,\\ 
 &\ {a}_2^\dagger\ket{\iota_{n_1,n_2;m_1,m_2}^\alpha}=\sqrt{n_2-m_2+1}\  \ket{\iota_{n+1,n_2+1;m_1,m_2}^\sigma}\ ,
\end{align}
\end{subequations}
with chosen $\iota_{n_1-1,n_2;m_1,m_2}^\beta, \iota_{n+1,n_2-1;m_1,m_2}^\delta \in {\cal I}_{v,n_1+n_2-1}^{\Gamma^{\cal A}}$, and $\iota_{n_1+1,n_2;m_1,m_2}^\gamma, \iota_{n+1,n_2+1;m_1,m_2}^\sigma \in {\cal I}_{v,n_1+n_2+1}^{\Gamma^{\cal A}}\ $. Equations \eqref{GenAC2} define the operators $\{a_i,{a}_i^\dagger\}$ using a choice of the mappings between the $G$ tensors $\iota$. These equations guarantee that
\begin{align}
 \forall i\in \{1,2\},\quad [a_i, {a_i}^\dagger]=\mathbb{I}_{{\cal H}_v^{\Gamma^{\cal A}}}\ .
\end{align}
However, the mappings between the $G$ tensors $\iota$ defining the actions of $a_1$ and $a_2$ are not independent, they are constrained by the condition
\begin{align}
 [a_1,a_2]=0\ .
\end{align}
Such consistent mappings exist but the choice is not unique. 
This means that there is a freedom in choosing the canonical operators $\{a_i,{a}_i^\dagger\}$, encoded in the choice of the $G$ tensors mapping. Hence one could adjust their choice of $\{a_i,{a}_i^\dagger\}$ to the coherence properties of the induced graph coherent states that one would like to recover for a preferred set of operators (observables). 
Since we are particularly interested in the Hamiltonian operator, an example of such consistent mappings is obtained by choosing the operators $\{a_i,{a}_i^\dagger\}$ as
\begin{align}\label{Exp1}
\forall i\in \{1,2\},\quad{\cal V}_i a_i&:=\text{Tr}_N^{(l)}\left[\tau^k h_{\alpha_i}\right]^\dagger\ ,\quad {\cal V}_i:=(a_i {a_i}^\dagger)^{-1/2}\ ,
\end{align}
similarly to \eqref{aToh}. Equations \eqref{Exp1} fix the choice of the mappings between $G$ tensors, for which one can express explicitly the matrix elements in a given basis, and they determine a set of annihilation and creation operators $\{a_i,{a}_i^\dagger\}$ satisfying \eqref{aCan}.

Coming back to the remark at the beginning of section \ref{S3} about the equivalent orientations for a pair of germs: if for instance the gauge group is $SU(2)$ and we choose the operator $\text{Tr}_N^{(l)}\left[\tau^k h_{\alpha}\right]$ to induce the graph change, we would have that two wedges $\omega_{KL}$ and $\omega_{LK}$ which have opposite orientation are in fact equivalent from the perspective of special loops, and the associated annihilation operators would not be independent. Explicitly, we would choose
\begin{align}
{\cal V}_{KL} a_{KL}&:=\text{Tr}_N^{(l)}\left[\tau^k h_{\alpha_{KL}}\right]^\dagger\ ,\\
{\cal V}_{LK} a_{LK}&:=\text{Tr}_N^{(l)}\left[\tau^k h_{\alpha_{LK}}\right]^\dagger\ ,
\end{align}
however, for $SU(2)$ we have that
\begin{align}
\text{Tr}_N^{(l)}\left[\tau^k h_{\alpha_{KL}}\right]^\dagger=- \text{Tr}_N^{(l)}\left[\tau^k h_{\alpha_{LK}}\right]^\dagger\ .
\end{align}
This means that in this case the two orientations are not independent, because the change of orientation generates a simple multiplication by a phase. Considering both wedges would induce a redundancy in the construction of the canonical structure at the vertex. Therefore, only one orientation should be considered when associating a canonical structure to such pair of germs $([e_K][e_L])$, and the second orientation would correspond to a different but not independent choice of canonical structure at the wedge.

Generalizing to an arbitrary number of wedges at the vertex $v$ is straightforward: given any state $\ket{\iota_{\{n_i\};\{m_i\}}^\alpha}$ ($i \in {\cal W}_v$) in ${\cal H}_v^{\Gamma^{\cal A}}$, equations \eqref{GenAC2} become
\begin{subequations}\label{GenAC}
\begin{align}
\forall j\in {\cal W}_v,\ a_j\ket{\iota_{\{n_i\};\{m_i\}}^\alpha} &=\sqrt{n_j-m_j}\  \ket{\iota_{\{n_1,\dots,n_j-1,\dots,n_{w_v}\};\{m_i\}}^\beta}\ ,\\ 
 {a}_j^\dagger \ket{\iota_{\{n_i\};\{m_i\}}^\alpha} &=\sqrt{n_j-m_j+1}\ \ket{\iota_{\{n_1,\dots,n_j+1,\dots,n_{w_v}\};\{m_i\}}^\gamma}\ ,
\end{align}
\end{subequations}
with a consistent choice of mappings between the $G$ tensors $\iota$. We call any pair of annihilation and creation operators $\{a_i,{a}_i^\dagger\}$ satisfying \eqref{OpInt1} and \eqref{GenAC} {\it generalized annihilation and creation operators}. Also, a complete set of generalized annihilation and creation operators, that is to each wedge of the vertex $v$ is associated a pair of generalized annihilation and creation operators, will be called a consistent canonical structure at $v$. The example \eqref{aToh} in Maxwell theory presented earlier corresponds to a (gauge invariant) choice of consistent canonical structure, while the choice in \eqref{Exp1} is an example of a (non gauge invariant) consistent canonical structure in the non Abelian case. Another example is given in section \ref{4a}.

\subsubsection{Graph coherent states in non-Abelian theories}\label{B2}
% \ \\
As pointed out in \eqref{Ker}, given a consistent canonical structure at a vertex $v$, the kernels of generalized annihilation operators are separable infinite dimensional Hilbert spaces and they do not coincide, similarly to the case of a multi-dimensional harmonic oscillator. However, unlike the harmonic oscillator, due to the presence of the tensorial structure at the vertices, one does not have a unique vacuum state. Namely, using the above notation, the solution to the system
\begin{align}\label{Vac}
 \forall i\in {\cal W}_v,\quad a_i \ket{\iota_{\{n_i\};\{m_i\}}^\alpha} = 0\ ,
\end{align}
is not unique. As a consequence of \eqref{GenAC2}, such states have necessarily $\{n_i\}=\{m_i\}$ and can simply be denoted $\ket{0_{v,\{m_i\}}^\alpha}$. By construction the states with no special loops satisfy \eqref{Vac}, but also for each distribution of special loops at a vertex, there exists at least one $G$ tensor such that the corresponding state satisfies \eqref{Vac}. Given a vacuum state at $v$, one can generate a subspace of ${\cal H}_v^{\Gamma^{\cal A}}$ by successive action of generalized creation operators, which is isomorphic to the space of a multi-dimensional quantum harmonic oscillator. To obtain the entire space ${\cal H}_v^{\Gamma^{\cal A}}$ one has to sum all the vector spaces generated from all vacuum states selected by the chosen canonical structure. We denote the space of these vacuum states ${\cal K}_v(\{a_i\})$.

Naturally, each choice of consistent canonical structure produces an overcomplete set of coherent states, the {\it generalized graph coherent vertices}. These states $\{\ket{Z_v}\}_{v\in \Gamma^{\cal A}}$ are defined as eigenvectors of the generalized annihilation operators
\begin{align}
 \forall v\in \Gamma^{\cal A},\quad \forall i\in {\cal W}_v,\quad a_i\ket{Z_v}= z_i\ket{Z_v}\quad,\quad Z_v:=\{ z_i\}\in\mathbb{C}^{w_v} \ ,
\end{align}
and can be obtained from the vacuum states selected by the canonical structure
\begin{align}\label{GGCV}
 \forall \ket{Z_v},\quad \exists!\ \ket{0_{v,\{m_i\}}^\alpha} \in {\cal K}_v(\{a_i\})\ :\quad \ket{Z_v} = \prod \limits_{i=1}^{w_v}e^{ z_i a_i^\dagger - \bar z_i a_i} \ket{0_{v,\{m_i\}}^\alpha} \ .
\end{align}
Hence we denote them as $\ket{Z_v, \{0_{v,\{m_i\}}^\alpha\}}$. Therefore the {\it generalized graph coherent states} are obtained as
\begin{align}\label{GGCS}
\ket{Z_{\Gamma^{\cal A}}, 0_{\Gamma^{\cal A}}}:= \bigotimes_{v\in \Gamma^{\cal A}}\ket{Z_v, \{0_{v,\{m_i\}}^\alpha\}}\ ,
\end{align}
they are labeled by a colored ancestor graph $\Gamma^{\cal A}$, a set of complex numbers $Z_v$ at each vertex of $\Gamma^{\cal A}$, these are the eigenvalues of the generalized annihilation operators, and a set of selected vacuum states, one at each vertex.

Taking the example in \eqref{Exp1}, one finds that, for the generalized graph coherent states obtained with this choice of canonical structure, the results for the variance and the relative variance computed in Maxwell case (from \eqref{Var0} to \eqref{RVar2}) hold in the non Abelian case, independently of the gauge group $G$ and the choice of vacuum states.

In the following section we discuss some aspects of the construction and its possible extensions and ramifications.

% \clearpage

\section{Discussion}\label{S4}

\paragraph*{a. Gauge invariance:}\label{4a}
\ \\
In the general construction we presented in section \ref{2B}, the canonical structure is obtained in the non gauge invariant Hilbert space ${\cal H}^{\text{vtx}}$. Also, in the example given by \eqref{Exp1}, the generalized annihilation operators are not gauge invariant, therefore the graph coherent states they induce would not be gauge invariant. There is however no obstacle in building $G$ gauge invariant coherent states, one has to simply choose a gauge invariant consistent canonical structure and restrict the $G$ tensors to intertwiners (gauge invariant tensors). An example of such structure is obtained by defining the generalized annihilation operators as
\begin{align}
\forall i\in {\cal W}_v,\quad {\cal V}_i a_i&:=\text{Tr}_N^{(l)}\left[h_{\alpha_i}\right]^\dagger \ .
\end{align}
Such choice provides gauge invariant vacuum states and one can use \eqref{GGCV} and \eqref{GGCS} to generate the gauge invariant graph coherent states.\\

\paragraph*{b. Gravity \& Yang-Mills:}
\ \\
As we mentioned earlier, the construction is independent of the dynamics of the theory. The graph change considered is indeed the one dictated by the dynamics of the polymer Yang-Mills as well as Einstein gravity, however the construction does not refer to any dynamics, only a specific graph change. Therefore the graph coherent states can be obtained in the context of any connection theory with a compact gauge group quantized following the polymer quantization, such as Yang-Mills and Einstein gravity. However, since gravity Hamiltonian combines holonomies and fluxes in a non linear fashion \eqref{GravHam}, we expect that one has to make a more elaborate choice of consistent canonical structure, other than \eqref{Exp1}, in order to induce coherence properties of the Hamiltonian operator of gravity with respect to the generated graph coherent states. This will be investigated in future works.\\

\paragraph*{c. Beyond the special loops:}
\ \\
The construction of graph coherent states presented in this article is tied to the graph change induced by the Hamiltonian operators \eqref{GravHam} and \eqref{YMHam} with the special loop regularization. This particular graph change allowed the decomposition of the Hilbert space ${\cal H}^{\text{vtx}}$ into stable separable subspaces, each with a structure which mimics the one of a multi-dimensional quantum harmonic oscillator. The additional tensorial structure at the vertices of a colored graph is manifest through the $G$ tensors mappings and the presence of infinitely many, but countable, vacuum states. However, this construction can in principle be extended to other graph changes. The main property that a graph change should abide, in order to realize the construction of graph coherent states, is to be able to induce a decomposition of the Hilbert space into stable separable subspaces, such that each of them is isomorphic to a space of a multi-dimensional quantum harmonic oscillator.\\

\paragraph*{d. Graph coherent states \& coarse graining:}
\ \\
In \cite{LSN} the authors introduce new states called the loopy spin ($SU(2)$-colored) networks. These states represent a coarse graining of the spin network states through enriching the tensorial structure at each vertex with abstract closed loops attached to it. The structure of the space of such coarse states, in the case of a bosonic statistics for the loops, is very similar to the structure of the spaces ${\cal H}_v^{\Gamma^{\cal A}}$ defined above. However, in our case the space is still the Hilbert space of the full quantum theory and the loops are associated to the wedges of a graph. The canonical structure that we defined is, though similar, very different from the one induced on the space of loopy spin networks. Hence the graph coherent states that we introduced take the form of different basis states superposition, and carry an entirely different interpretation than the one of a canonical coherent state on the space of loopy networks. Nevertheless, the similarities between the structures of the spaces suggest that there is perhaps a possibility to join the two frameworks in the direction of inducing a coarse grained dynamics for the loopy spin networks. This is to be investigated in future works.

\section{Summary \& outlooks}

In this article, we introduced a new family of coherent states on the Hilbert space of a polymer quantum theory of connections with an arbitrary compact gauge group $G$. These states take the form of a superposition of basis network states with different graphs, hence the appellation {\it graph coherent states}. Inspired from the quantum dynamics of Yang-Mills and gravity, the notion of a graph change lies at the core of the construction. 
This one starts by introducing a consistent canonical structure on the stable subspaces of the graph change, generating a Fock-like structure similar to that of a multi-dimensional quantum harmonic oscillator, but with additional degrees of freedom that are the $G$ tensors (or intertwiners on the gauge invariant space). The canonical structure consists of generalized canonical annihilation and creation operators, which encode mappings between $G$ tensors and a set of vacuum states. The graph coherent states are then given as eigenvectors of the generalized annihilation operators. A set of complete coherent states is not unique, as there is some freedom in the choice of the canonical structure encoded in the choice of consistent mappings between $G$ tensors.

In addition to the standard coherence with respect to the canonical operators, particular graph coherent states can exhibit coherence properties with respect to operators inducing the specific graph change, such as the ones involved in the Hamiltonian operators in Yang-Mills or gravity theories, but also with respect to operators depending only on fluxes, when these describe a more global information concerning the graphs (e.g.\ the operator in \eqref{GermFlux}). Taking into account the freedom in the choice of the canonical structure, one could build graph coherent states more adapted to the operators that one would like to investigate. In particular one could hope to gain more insight about the behavior of the quantum dynamics with graph changing Hamiltonian operators, since the spectrum of such operators is yet inaccessible.

Finally, there are many avenues which can be explored in the context of those graph coherent states. Namely, the possible link to a notion of coarse graining, e.g.\ the loopy spin networks \cite{LSN}; the realization of the construction for other graph changes; the derivation of an effective dynamics in a certain sector captured by a subset of the graph coherent states; and a new perspective on the semi-classical limit and the continuum limit. We leave these questions for future research.
% \clearpage

\section*{Acknowledgements}
The author thanks Ilkka M\"{a}kinen, Jerzy Lewandowski, Etera Livine, Christian Fleischhack and Daniel Siemssen for fruitful discussions, clarifications and comments. This work was supported by the grant 2011/02/A/ST2/00300 of the Polish Narodowe Centrum Nauki (NCN) and the project BA 4966/1-1 of the German Research Foundation (DFG).

% \clearpage

\bibliography{references}

\end{document}